# Textual analysis of clinical notes on pathology request forms to determine sensitivity and specificity of Hepatitis B and C virus infection status

Eric H. Kim[1], Brett A. Lidbury[1], Alice M. Richardson[1,2*]

[1]National Centre for Epidemiology and Population Health, Australian National University, Canberra ACT 2601, Australia

[2]Statistical Support Network, Australian National University, Canberra ACT 2601, Australia

*Correspondence: alice.richardson@anu.edu.au

**ABSTRACT:** 180 words (max 250 words)

*Background*: It is not established whether clinical notes provided on pathology request forms are useful as decision support data when assessing Hepatitis B and C viral infection status.

*Objective*: To determine sensitivity, specificity, and predictive value of clinical notes for identifying infection status of Hepatitis B and C.

*Methods*: The study comprises 179 cases and 166 cases tested for HBsAg and anti-HCV serological markers, respectively, and accompanied by a written description (clinical note) provided on pathology request forms by the clinician on duty. The clinical note sensitivity, specificity, positive (PPV) and negative (NPV) predictive values were calculated using serological HBsAg and anti-HCV tests as gold standards.

*Results*: The sensitivity and specificity of clinical notes for Hepatitis B infection status were 90% and 56%, respectively. The sensitivity and specificity of clinical notes for Hepatitis C infection status were 86% and 21%, respectively.

*Conclusions*: Clinical note information identifies moderate-to-high sensitivity with regards to Hepatitis B and C viral infection status, however, given low specificity in both groups, the clinical note is not favourable for ruling disease "in", possibly due to high rate of false positives.

# INTRODUCTION

Chronic hepatitis B and C infections are a major public health problem associated with significant morbidity and mortality worldwide. Globally, an estimated 257 and 71 million people are living with chronic hepatitis B and C infection, respectively [1]. Without intervention, about 15-40% of chronically infected individuals can lead to development of liver cirrhosis, end-stage liver disease (ESLD) or hepatocellular carcinoma (HCC), and in severe cases, require liver transplantation [2-4].

Clinical detection is based, in part, on recognition of the multiple clinical signs and symptoms. These clinical histories are typically documented; however, a conclusive diagnosis is not necessarily synthesised by the primary care physicians until more specific diagnostic tests are completed to support a definitive diagnosis. Importantly, clinical notes are often provided by clinicians on pathology laboratory request forms, without consideration of the uses to which they could be put in secondary analyses. The availability of computing technology to extract, read and mine text such as these notes is at the same time increasing.

Clinical notes play a critical role for communication providing synthesis, summary of information, and decision making involved in patient management [9-11]. A large portion of clinical findings mentioned in clinical reports (e.g., discharge summaries, pathology reports) contain discursive references about the disease in general. In particular, narrative reports such as pathology reports convey valuable diagnostic information that is predictive of the prognosis and biological behaviour of a disease process [12]. While the use of large repositories of patient-specific clinical reports [5] and routine pathology data [6-8] has been assessed for their predictive prognosis value, it is not yet established whether clinical notes provided on pathology request forms provide a good predictor for disease outcome.

The clinical note on pathology test request forms is a section typically containing clinical information about the patient, and a description of possible diagnosis prior to the outcome of pathological tests. Currently, free-text reporting is the norm. Although the clinical note is not intended for documenting extensive information, it can provide additional information pertinent to the clinician who index them. The information may contain negation statements to support a diagnosis, quantified statements such as words "all" or "some", or contain a series of queries that is obscure from histories and requires a follow-up. This information, in conjunction with pathology test outcome provide prognostic value and are used by clinicians to support decision making on appropriate treatment.

The interpretation of pathology tests and accompanying clinical notes, particularly those involved with screening, hence is intricate and relies on an understanding of the diagnostic accuracy of the test and the prevalence in the community of diseases that the test can predict.

The objective of the current study is therefore to understand the diagnostic or prognostic value of clinical notes that accompany inpatient pathology lab requests, for the diagnosis of HBV or HCV infection status. We hypothesise that the clinical notes provide useful support data to suggest status of infection and improves diagnostic decision-making outcome. More specifically, we investigated the sensitivity and specificity of clinical notes, which is calculated based on serological HBsAg and anti-HCV antibody tests as gold standards for Hepatitis viral infection status.

**MATERIALS AND METHODS**

**Research ethics**

For access to de-identified patient data, human ethics for this study was approved by The Australian National University Human Ethics Committee (protocol no. 2012/349) and the ACT Health Human Research Ethics Committee (ETHLR.11.016).

**Clinical note**

In the present study, the "clinical note" accompanies a pathology request, in which the clinician denotes relevant patient information based on history and physical examination prior to serological testing. The clinical note is recorded by free text entry. The style of note adopted is by physician preference that best fits their workflow, and the interpretation is restricted to the individual physician or lab technician on duty.

**Data set and text mining**

The data set used in this study originally comprised 18624 individuals tested for hepatitis virus in the period 1997 – 2007. The data was provided by ACT Pathology, The Canberra Hospital (TCH), Australia, which has been described and analyses elsewhere [7, 8].

Pathology records were assessed, and manual inspection of records and spreadsheet sorting was initially performed by one of the authors (BAL) to divide clinical notes for all 18624 subjects into 46 subjective categories based on disease and health conditions (**Table S1**). A second author (EK) refined the categorisation based on implementing the International Classification of Diseases version 10 (ICD-10) codes by World Health Organization (WHO) [13], which served as a primary guideline for categorisation of clinical notes used in this study. We retrospectively selected observations with information on serological HBsAg, anti-HCV antibody tests, and clinical notes available on pathology request.

Of the 46 categories (**Table S1**), we have defined 2 categories that belong to suspected Hepatitis B and C infection status. For example, as outlined on **Table S1**, clinical note "Hepatitis B positive" is assigned as Category 1, and "Hepatitis C positive" as Category 2. In addition, we have investigated sensitivity and specificity in 10 additional clinical code categories, which were selected to serve as controls. The selection is based subjectively on clinical note categories with adequate number of cases and availability of hepatitis virology data, and representing various disease states that may or may not affect risk of hepatitis. These categories comprise leukopenia, psychological illness, alcohol abuse, gastrointestinal disorders, liver diseases, rheumatological disorders, pregnancy screening, intravenous drug abuse, work pre-screening, and fatigue (Table S1 shaded rows). Results for HBV vaccination status were removed prior to analysis.

**Reference standards and prespecified cut-off points**

A description of variables used in this study is shown in **Table 1**. For diagnostic evaluation of each clinical note, i.e., sensitivity, specificity, PPV, NPV, LR+, LR- [14-17], serum HBsAg and anti-HCV values were used as reference gold standards for comparison. Serum HBsAg was classified as positive at $\geq 1.6$ immunoassay units (IU) and anti-HCV antibody $\geq 1.0$ IU for a positive classification. All other HBsAg and anti-HCV antibody results below this assay cut-off were classified as negative (M. De Souza, ACT Pathology, pers. comm.). Additional data cleaning included the removal of subjects with missing serological values.

**Software and analysis**

All descriptive statistical analyses and figures were produced using SPSS statistics version 26 (IBM) and Origin (OriginLab) software, respectively.

The 2 x 2 contingency tables were constructed to calculate sensitivity and specificity, which are defined in the usual way [14-17]. A working numerical example is illustrated in **Figure S1**.

**RESULTS**

**Demographic and clinical data of subjects**

A retrospective analysis was performed on collected clinical data from 1997-2007. Demographic and clinical characteristics of these subjects are shown in **Figure 1** and **Table 2**. Among these categories, we specifically aimed at investigating the predictive nature which described status of Hepatitis B and C infections (**Table S1**; categories 1 and 2, respectively). In the Hepatitis B group (category 1), 179 cases were analysed from a total of 241 cases, with missing serological data of 62 cases which was excluded (**Table 2**). The mean age was 38 (SD 14.4) years, with M:F ratio of 98:81 cases. In the Hepatitis C group (category 2), 166 cases were analysed from a total of 337 cases, with missing serological data of 161 cases which was excluded (**Table 2**). The mean age was 36 (SD 15.8) years, with M:F ratio of 85:81 cases.

**Diagnostic predictability of key clinical notes for Hepatitis B**

To explore the diagnostic suitability of clinical notes that accompany pathology request for Hepatitis B as the standard for a correct diagnosis, the sensitivity and specificity of clinical note (Category 1) was calculated and compared to HBsAg serology as true gold standard in the 179 subjects (**Table 3** and **Figure S1**).

The sensitivity and specificity values are 0.90 and 0.56, respectively. This indicates that clinical note for Hepatitis B is appropriate for ruling disease "out", given high sensitivity value, however, not favourable for ruling disease "in", given moderate specificity of the results, i.e., there is risk of high false-positive results. The positive (PPV) and negative (NPV) predictive values are 0.61 and 0.87, respectively.

In addition, the positive (LR+) and negative (LR-) likelihood ratios are 2.05 (95% CI: 1.61-2.56) and 0.18 (0.09-0.37), respectively are both in the range of a small-to-moderate effect, i.e., there is a 2-fold increase in the odds of HBV infection state, if the clinical note identifies as positive HBV, and a 5-6 fold decrease in the odds of HBV infection state, if the clinical note identifies as negative HBV [16, 17].

**Diagnostic predictability of key clinical notes for Hepatitis C**

To explore the diagnostic suitability of clinical notes that accompany pathology request for Hepatitis C as the standard for a correct diagnosis, the sensitivity and specificity of clinical note (Category 2 Hepatitis C) was calculated and compared to anti-HCV antibody serology as the true gold standard in the 166 subjects (**Table 3** and **Figure S1**).

The sensitivity and specificity values are 0.86 and 0.21, respectively, suggesting moderate sensitivity, however, low-to-poor specificity. This indicates that clinical note for Hepatitis C status is not a reliable tool for diagnosis. The positive (PPV) and negative (NPV) predictive values are 0.73 and 0.37, respectively.

In addition, the positive (LR+) and negative (LR-) likelihood ratios are 1.08 (95% CI: 0.92-1.27) and 0.67 (0.34-1.40), respectively is in the range of low-to-moderate effect, suggesting clinical notes provided for Hepatitis C would have a neutral or no effect for influencing post-test probability.

**Diagnostic predictability of hepatitis B and C in other clinical notes**

We additionally evaluated diagnostic predictability of hepatitis B or C infection status in further ten clinical notes, which represent a wide varying disease states (**Figures 2 and 3; Tables 4 and 5**). The first aim was to verify specificity, i.e., hepatitis note should not assign to other unrelated infection or disorders, and the ten clinical notes serve as negative controls.

Second, our hypothesis was that in some categories with higher risk of hepatitis virus infection, for example, liver diseases (category 24) or fatigue (category 37), we would observe higher sensitivity (possibility of prior infection) compared to categories with lower risk, for example, gastrointestinal disorders (category 22), and serve as surrogate positive controls.

Interestingly, in terms of assessing Hepatitis B infection status, the sensitivity (**Figure 2A; Table 4**) and specificity (**Figure 2B; Table 4**) of ten clinical notes showed low sensitivity (<0.17) and high specificity (>0.98) across all ten categories. This suggests Hepatitis B infection status is poorly detected in other clinical note categories (low sensitivity), however, high specificity was obtained.

In terms of assessing Hepatitis C infection status, the sensitivity (**Figure 3A; Table 5**) and specificity (**Figure 3B; Table 5**) of ten clinical notes also showed low sensitivity (<0.11) and high specificity (>0.99) across all ten categories. This suggests Hepatitis C infection status is poorly identified in other clinical note categories (low sensitivity), however, high specificity was obtained.

Overall, given low sensitivity, these results support that clinical notes are not a suitable standard approach for evaluating diagnostic capacity of Hepatitis B or C infection status across other varying disease states, and that clinical note does not discriminate the status of Hepatitis B or C infection based on its prior health risk.

However, the high specificity in both tests suggest its utility for detecting a true negative. For instance, work pre-screen (category 32) captures all persons presenting for a screening test with no symptoms or pre-existing conditions. This group acts as a natural control group in the experiment, who would be expected to have low likelihood for presence of hepatitis. In this group, the data showed 100% of persons were negative for hepatitis infection status (i.e., specificity 1.00 for both Hepatitis B and C), suggesting its applicability for capturing true negatives.

## DISCUSSION

The objective of the current study was to understand the predictive power for the use of clinical notes accompanied by inpatient pathology lab requests in the diagnosis of hepatitis infection. Specifically, we investigated the sensitivity and specificity of clinical notes, based on serological HBsAg and anti-HCV antibody tests as gold standards for Hepatitis viral infection status.

**Hepatitis B, C findings:** The results of our study demonstrate that the sensitivity of clinical notes for both Hepatitis B and C status show moderate-to-high values (90% and 86%, respectively), which suggests that written clinical notes provided at the time of pathology request display good sensitivity based on clinical history and individual clinician assessment for the diagnosis of HBV and HCV infection status. The calculated specificity for both clinical notes, however, show low values (56% and 21%, respectively), which suggest weak performance for identifying HBV and HCV infection outcomes, and incorrectly identifying patients who do not have the condition.

Compared to literature, the current diagnostic serological tests for detection of Hepatitis infections show an overall a high level of diagnostic accuracy with sensitivity and specificity for HBV (96-99%) and HCV (98%) [18-20]. Hence, the preliminary findings in the current study suggest that clinical notes are at best moderately useful in the identification of patients with Hepatitis B and C infections (moderate-to-good sensitivity), however not useful to be employed as a sole source of diagnosis of Hepatitis infection status (low specificity), and require further information and confirmation with other tests.

Also, likelihood ratios can be calculated in order to estimate the post-test probability of hepatitis infection status with a positive test compared to probability of infection status with a negative test. In our results, LR+ (2.05) and LR- (0.18) values for Hepatitis B, suggest a small-to-moderate effect for influencing post-test probability of Hepatitis B with a positive test compared to probability of Hepatitis B with a negative test. The LR+ (1.08) and LR- (0.67) values for Hepatitis C suggest that the predictive capability of clinical note usage for establishing Hepatitis C infection is limited [16, 17].

Additionally, attention should be paid to predictive values (PPV and NPV) which may be valuable to the physician in the clinical practice; with caution that predictive values are influenced substantially by disease prevalence [21]. This factor should also be considered in our current data set (1997 – 2007) which spans over ten years. The prevalence data for acute hepatitis B and C infections in Australia showed that it remained relatively steady from 1997 to 2007 (For Hepatitis B infection: 13618 affected persons in 1997 in comparison to 13906 affected persons in 2007) (**Figure S2**) [22]. Hence, if we consider steady prevalence in the given study time frame (1997-2007), our data suggest that clinical notes identified 61% of those with a positive clinical history, as having Hepatitis B infection status (PPV 0.61), while clinical notes identified 87% of those with a negative clinical history, unaffected by Hepatitis B (NPV 0.87). In comparison, clinical notes identified 73% of those with a positive clinical history, as having Hepatitis C infection status (PPV 0.73), while identifying 37% of those with a negative history, unaffected by Hepatitis C (NPV 0.37). The steady acute hepatitis B infection during 1997-2007 may be attributed to free infant vaccination as part of National Immunisation Program that was first introduced nationally in 1997 [23]. It is worth noting that the prevalence has since gradually increased to reach 16224 persons by year 2017.

**Other clinical note findings:** We evaluated individuals suspected of having an HBV and HCV infection in other clinical note categories. As illustrated in **Figures 2 and 3**, the sensitivity and specificity of ten clinical notes showed low sensitivity but high specificity across all ten categories. This suggests HBV and HCV infection status is poorly detected in other clinical note categories (low sensitivity), however, may provide a reliable tool for ruling disease "in" (high specificity), as indicated by high true negative rates [14, 15]. Diagnostic tests preferably require high levels of both sensitivity for detection of cases early in disease course, and specificity to minimise the risk of high false-positive results. Hence, these results overall support that clinical notes is not a suitable standard approach for evaluation of the diagnostic capacity of HBV or HCV status.

**Limitations and future directions of study**

In assessing the generalisability for the use of clinical notes as a diagnostic tool, there are several important limitations of the present study.

*Clinical note and disambiguation*: The complexity and descriptive nature of clinical notes provide several potential sources of error. Firstly, due to the qualitative nature of clinical notes, complications occasionally arose during assigning clinical notes to a "positive" or "negative" hepatitis virus infection state. For instance, there was difficulty interpreting the intention of written note provided by clinicians (intra-rater variability), and in addition, each individual clinician provided different script or note-taking style (inter-rater variability). We decided that when a "statement" is made on the written note, such as "Hep C", "Known Hep C", "Hep C Pos", "Hx Hep C", "Hep C exposure", it was considered a positive Hepatitis C infection status, whereas a written note with a "query" or non-specific data entry, such as "?Hep C", "Possible Hep C", "Hepatitis cause?", "Screen Hep C", was considered a negative infection status. Hence, style of written note provided by a different clinician, as well as interpretation of clinical notes by the interpreter, both provided sources of error to the study.

Despite these limitations, clinical note documents form core aspect of patient care and management. Among these, free text entry is typically considered crucial type of written communication between health providers for presenting complex sets of information. Clinicians appreciate the efficiency and flexibility of narrative free text [24], however, this method can be challenged by lengthy, medical terminology that is difficult to review, or potentially contain erroneous information that can be transferred between providers. Alternatively, the benefits of structured data entry include data reuse for downstream applications such as education, research, and quality assurance [25]. Despite this, the disadvantages of using structured data entry include inefficient data entry and retrieval, difficulty creating and using standardised computerised documentation system, and use should be considered that best fits practice [26, 27].

*Number of cases and testing*: Another risk of bias is limited by inadequate data. Despite a large total number of subjects in the study (n = 18624), text mining extracted available data to 179 cases for clinical note Hepatitis B category, and 166 cases in the Hepatitis C category. In addition, reduced number of cases was also due to lack of serological data available within each category, which was excluded at the outset of study. Lastly, the data was unspecific to which year serological testing was performed for each patient in our current study. An example of future study may include a case-control study with similar prevalence in a clinical setting, with data available on collection date of serology.

*Generation of cut-off points and ROC curve*: In general, to evaluate the diagnostic ability of a new test, empirical receiver-operating characteristic (ROC) curve is prepared and analysed for the optimisation of cut-off value where both measures of sensitivity and specificity provide the best results [14, 28]. However, in the present study, the ROC curve was unable to be generated as descriptive clinical notes were converted into categorical variables, i.e., allocated into a binary "positive" and "negative" Hepatitis virus infection status.

*Changes in viral load, seroconversion and seroclearance*: HBsAg is a hallmark of HBV infection, and one of the first serological marker to appear during acute infection, as well as chronic HBV infection for HBsAg persisting over 6 months. As such, serum HBsAg levels can fluctuate over time, which may require measurement of seroconversion [29] or sequential follow-up samples to potentially overcome this problem. Additionally, some studies have shown that serum HBsAg levels can decrease spontaneously over time, termed seroclearance, during the inactive phase of chronic HBV infection [30, 31] or for patients undergoing antiviral therapy [32]. The limitation of use of clinical notes for diagnostic tool is the difficulty for providing quantitative changes on viral load, however, may be useful for identifying qualitative aspects, such as high risk encounter or previous hepatitis history, associated with a high likelihood of infection status.

*Genetic differentiation of HBV and HCV*: Clinical note does not differentiate genetic variability of HBV and HCV, which can pose a major diagnostic challenge. Currently, more sensitive nucleic acid testing such as polymerase-chain reaction (PCR) for HBV-DNA or HCV-RNA is widely available for differentiating genotypes and quantifying viral load [33-35].

**CONCLUSION**

Hepatitis B and C infections are a significant global health burden leading to liver cirrhosis and end-stage liver disease without intervention. The current reliable diagnostic tool for identifying acute and chronic infection status rely on serology, virology, and nucleotide-based techniques. In our present study, we investigated whether written clinical notes provided on pathology request forms provide useful tool for providing information on Hepatitis virus infection status in patients. The preliminary findings suggest that clinical notes are moderately useful in the identification of patients with Hepatitis B and C infections, however not useful to be employed as a sole source of diagnosis of Hepatitis infection status, and require further information and confirmation with other tests.

**Funding information**

BAL was supported by funding granted by the Quality Use of Pathology Programme (QUPP), the Commonwealth Department of Health, Canberra Australia.

**Author contributions**

Conceptualization, BAL and AMR;
Methodology, BAL, AMR and EK;
Software, EK;
Validation, BAL, AMR;
Formal Analysis, EK;
Investigation, EK;
Resources, BAL and AMR;
Data Curation, BAL and EK;
Writing – Original Draft Preparation, EK;
Writing – Review & Editing, BAL, AMR and EK;
Visualization, EK;
Supervision, BAL and AMR;
Project Administration, AMR;
Funding Acquisition, BAL

**Conflict of Interest**

The authors declare no conflict of interest

**Ethics Statements**

The current study was approved by The Australian National University Human Ethics Committee (protocol no. 2012/349) and the ACT Health Human Research Ethics Committee (ETHLR.11.016), with additional human ethics approval also granted by ACT Health.

**Author Biography**

E. H. Kim is a medical student at the Australian National University. He has previously graduated with BSc (Hons), and PhD from the University of Auckland.

Brett A. Lidbury is an Associate Professor with the National Centre for Epidemiology and Population Health (NCEPH), The Research School of Population Health, the Australian National University, and a Visiting Associate Professor with the Department of Health Evidence, Radboud University Medical Centre, Nijmegen, The Netherlands. Educational attainment includes a B.Sc. (Hons) degree from the University of Newcastle and PhD from the Australian National University. Recent recognition was also granted via the award of a Fellowship to the Science Faculty of the Royal College of Pathologists of Australasia (RCPA). Research experience spans laboratory-based biomedicine, pathology, infectious diseases and the application of machine learning to biomedical investigations. Specific interests currently involve pathology quality and predictive modelling, as well as cross-disciplinary fundamental research into Myalgic Encephalomyelitis (ME). Educational contributions include the inception and teaching of courses in molecular biology, genetics and laboratory medicine.

Associate Professor Alice Richardson is Lead of the Statistical Support Network at the Australian National University (ANU). She graduated with BA(Hons) in Statistics and Operations Research from

Victoria University of Wellington, New Zealand, then MStats and PhD in Statistics from the ANU. She has twenty years of experience in teaching undergraduate statistics at the University of Canberra. From 2016 - 2019 she was biostatistician at the National Centre for Epidemiology & Population Health at the ANU. Her research interests are in linear models and robust statistics; applications of statistical methods to large biomedical data sets; and innovation in statistics education. She is a member of the Statistical Society of Australia (SSA), and the New Zealand Statistical Association, amongst others. In 2014 she received a Service Award from the SSA and in 2017 she was awarded a Visiting Fellowship from the Victorian Centre for Biostatistics, Melbourne.

# FIGURES AND TABLES

## Table 1. Description of variables

| Variable | Description and definition | Measurement units |
|---|---|---|
| **Reference standard:** | | |
| HBsAg | Hepatitis B Surface Antigen (marker of HBV infection) | Positive (>=1.6 IU) <br> Negative (< 1.6) |
| Anti-HCV antibody | Patient antibody to HCV (anti-HCV immunoglobulin G), indicating contact with virus | Positive (>= 1.0 IU) <br> Negative (< 1.0) |
| **New test: Clinical note** | | |
| Clinical note | Initially grouped into 46 categories, based on disease and health conditions | see Table S1 |
| **Explanatory variable:** | | |
| Age | Age of subject | Years |
| Sex | 1 = M; 2 = F | M or F |
| **Evaluation parameters:** | | |
| Sensitivity (Sn) | Probability that when disease is present, the test is positive | |
| Specificity (Sp) | Probability that when disease is absent, the test is negative | |
| Positive predictive value (PPV) | Probability that a person who has a positive test, actually has the disease | |
| Negative predictive value (NPV) | Probability that a person with a negative test, actually does not have the disease | |
| Positive likelihood ratio (LR+) | How many times more likely a person has disease, when you have a positive test | |
| Negative likelihood ratio (LR-) | How many times less likely a person has disease, when you have a negative test | |

**Table 2. Demographic and clinical characteristics of the subjects**

|  | All subjects | Clinical note: Hepatitis B | Clinical note: Hepatitis C |
|---|---|---|---|
| **Number (*missing data)** | 18624 | 179 (62) | 166 (161) |
| **Age (mean years ± S.D.)** | 40 ± 17.6 | 38 ± 14.4 | 36 ± 15.8 |
| **Gender** | Male (n = 7539) | Male (n = 98) | Male (n = 85) |
|  | Female (n = 11076) | Female (n = 81) | Female (n = 81) |

*Missing data: serological data not available or unspecified

**Table 3. Predictive validity of clinical notes for Hepatitis B and C**

| Clinical note | Sn (%) (95% CI) | Sp (%) | PPV (%) | NPV (%) | LR+ | LR- |
|---|---|---|---|---|---|---|
| **Category 1: Hepatitis B** | 90 (80.6-95.4) | 56 (45.7-65.7) | 61 (54.9-65.9) | 87 (78.3-93.4) | 2.05 (1.61-2.56) | 0.18 (0.09-0.37) |
| **Category 2: Hepatitis C** | 86 (77.9-91.4) | 21 (10.5-35.0) | 73 (69.3-75.8) | 37 (22.5-54.4) | 1.08 (0.92-1.27) | 0.67 (0.34-1.40) |

Sn, sensitivity, Sp, specificity, PPV, positive predictive value; NPV, negative predictive value; LR+, positive likelihood ratio; LR-, negative likelihood ratio.

**Table 4. Percentage of HBsAg positive results according to ICD-10 Code (See Table S1 for ICD and subjective classification of notes recorded on pathology requests).**

| ICD-10 Code | Code Summary | Percentage (%) HBsAg Positive | Sn (%) | Sp (%) | PPV (%) | NPV (%) |
|---|---|---|---|---|---|---|
| I | Primary HBV suspected | 43 | 90 | 56 | 61 | 87 |
| III | Leukopaenia: | 0.9 | 0 | 100 | n.d. | 99 |
| V | Psychological illness or psychosis | 2.8 | 0 | 100 | n.d. | 97.2 |
| V | Alcohol abuse | 0 | n.d. | 100 | n.d. | 100 |
| XI | Gastrointestinal disorders: IBD, IBS | 0 | n.d. | 100 | n.d. | 100 |
| XI | Liver diseases and deranged LFTs | 0.8 | 0 | 97.7 | 0 | 99.2 |
| XII and XIII | Rheumatological disorders | 0.7 | 0 | 99.6 | 0 | 99.3 |
| XV | Pre- and post-pregnancy screening | 0.8 | 17.2 | 99 | 71.4 | 99 |
| XX | IVDA | 1.9 | 0 | 99 | 0 | 98 |
| XXI | Work screening | 0 | n.d. | 100 | n.d. | 100 |
| - | Fatigue and lethargy | 2.2 | 0 | 100 | n.d. | 97.8 |

n.d. = not defined

**Table 5. Percentage of anti-HCV positive results according to ICD-10 Code (See Table S1 for ICD and subjective classification of notes recorded on pathology requests).**

| ICD-10 Code | Code Summary | Percentage (%) anti-HCV Positive | Sn (%) | Sp (%) | PPV (%) | NPV (%) |
|---|---|---|---|---|---|---|
| I | Primary HCV suspected | 71 | 86 | 21 | 73 | 37 |
| III | Leukopaenia | 0 | n.d. | 100 | n.d. | 100 |
| V | Psychological illness or psychosis | 17 | 0 | 100 | n.d. | 83 |
| V | Alcohol abuse | 12.4 | 0 | 99 | 0 | 88 |
| XI | Gastrointestinal disorders: IBD, IBS | 0 | n.d. | 100 | n.d. | 100 |
| XI | Liver diseases and deranged LFTs | 3.9 | 11.1 | 99.3 | 40 | 96.1 |
| XII and XIII | Rheumatological disorders | 3.0 | 0 | 100 | n.d. | 97 |
| XV | Pre- and post-pregnancy screening | 2.5 | 5.1 | 100 | 100 | 97 |
| XX | IVDA | 46 | 1.3 | 100 | 100 | 54 |
| XXI | Work screening | 8.5 | 0 | 100 | n.d. | 91.5 |
| - | Fatigue and lethargy | 0.6 | 0 | 100 | n.d. | 99.4 |

n.d. = not defined

**Figure 1. Patient demographic data used in this study**

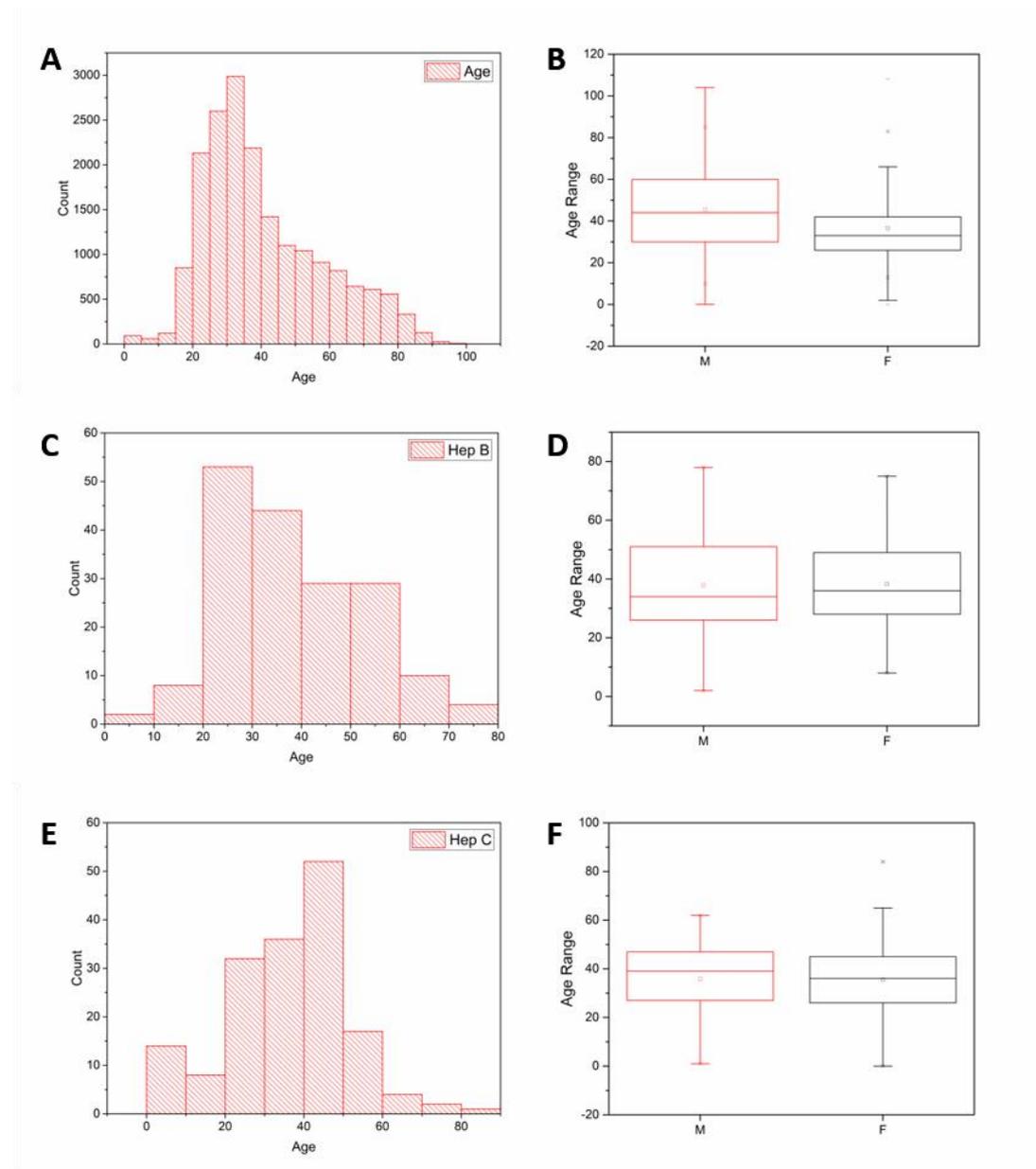

Age and sex distribution. A-B illustrates (A) age and (B) sex distribution of all subjects in the study (n = 18624); C-D illustrates (C) age and (D) sex distribution in the clinical note Hepatitis B group (n = 179); E-F illustrates (E) age and (F) sex distribution in the clinical note Hepatitis C group (n = 166).

**Figure 2. Comparison of sensitivity and specificity between other clinical notes and Hepatitis B**

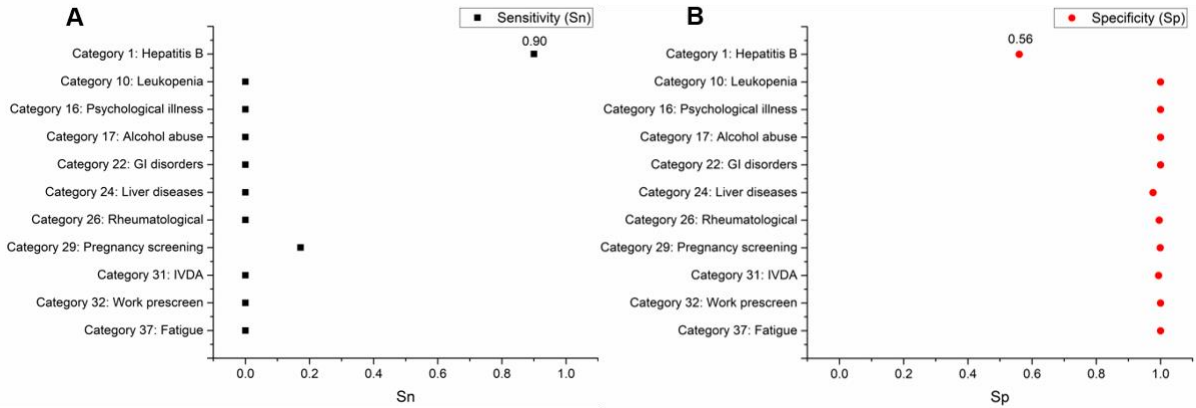

(A) Sensitivity and (B) Specificity in ten other clinical notes. Low sensitivity (<0.17) and high specificity (>0.98) detected across all other clinical notes, compared to Hepatitis B infection status (Sn and Sp, 0.90 and 0.56, respectively).

**Figure 3. Comparison of sensitivity and specificity between other clinical notes and Hepatitis C**

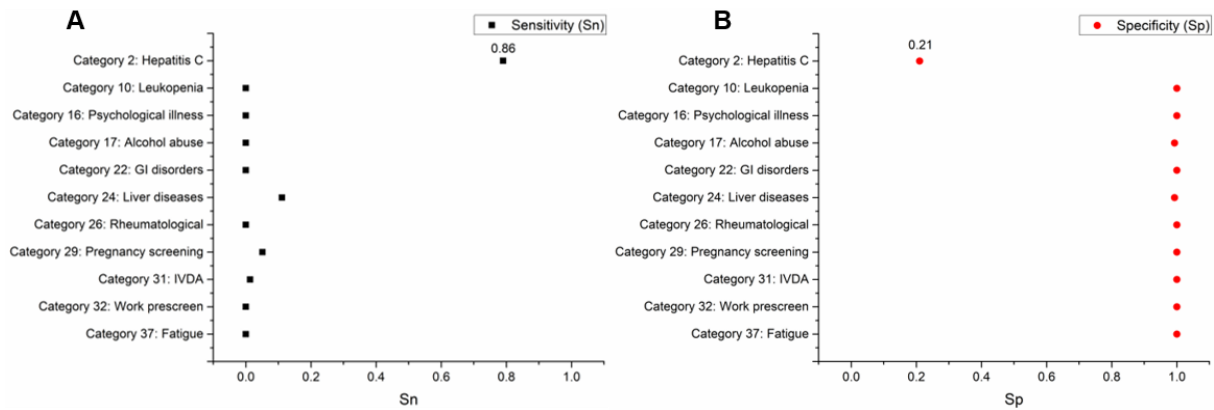

(A) Sensitivity and (B) Specificity in ten other clinical notes. Low sensitivity (<0.11) and high specificity (>0.99) detected across all other clinical notes, compared to Hepatitis C infection status (Sn and Sp, 0.86 and 0.21, respectively).

# SUPPLEMENTARY MATERIAL

**Table S1. Summary of clinical notes** (**shaded areas* – denotes clinical note categories analysed)

| ICD-10 nomenclature | Category | Number of subjects (n) | Summary of clinical notes on pathology form |
|---|---|---|---|
| I (B15-B17) | 1* | 241 | Hepatitis B |
| I (B15-B17) | 2* | 327 | Hepatitis C |
| I | 3 | 124 | Hepatitis (not categorised as A, B or C), AIH ("active hepatitis", "autoimmune hepatitis") |
| I | 4 | 153 | Infection and Inflammation: Fever or febrile, pyrexia, sepsis, septic shock, sweats, raised WCC (WBC), PUO (pyrexia of unknown [uncertain] origin), allergy (i.e., raised eosinophils), CRP (c-reactive protein), uveitis, asthma, swollen glands, nodes |
| I | 5 | 70 | HIV positive, ARVT/ART (anti-retroviral treatment) |
| II | 6 | 112 | Tumour: Carcinoma, ascites (AS) formation, SCC (small cell carcinoma), chemotherapy, POA (pancreatic onco-foetal antigen) |
| II | 7 | 108 | Lymphoma: NHL ("Non-Hodgkin's Lymphoma"), Hodgkin's disease |
| II | 8 | 198 | Leukaemia and Lymphoma: Leukaemia (CLL, CML, AML, ALL, APML), lymphoma, lymphadenopathy, lymphocytosis, lymphoproliferative disorders, myeloma, multiple myeloma (Bence-Jones proteins, IgG paraproteins), splenomegaly, chemo/pre-chemo |
| III | 9 | 200 | Red cell counts: Anaemia (microcytic, macrocytic, pernicious, B12, ferritin), haemochromatosis, H.H (hereditary haemochromatosis), polycythaemia, porphyria, haemolytic disease, transfusion reactions, spherocytosis, thalassaemia, sickle cell anaemia, G6PD deficiency, haem- disorders (haemoptysis), hyperferritinaemia/ferritin |
| III | 10* | 135 | Low white counts: Neutropaenia, lymphopaenia, agammaglobulinaemia |
| III | 11 | 285 | Blood clotting and vascular disorders: Thrombocytopaenia, low platelets, bleeding, on warfarin, INR, thrombocytosis, pancytopaenia, DVT, purpura, bruising, haemophilia, ITP (immune thrombocytopaenic purpura), ischaemia, vasculitis, TTP (Thrombotic Thrombocytopaenic Purpura), abnormal bleeding, Leiden (Fva) mutation, Von Willebrands (VWD), thrombosis, PEA/PE (pulmonary embolus) |
| IV | 12 | 57 | Thyroid problems ("on thyroxine", "oroxine"), goitre, Grave's disease |
| IV | 13 | 22 | T1DM (Type-1 diabetes mellitus) |
| IV | 14 | 112 | T2DM (Type-2 diabetes mellitus) |
| IV | 15 | 72 | Nutrition: Weight loss, poor diet, malnutrition, anorexia, "failure to thrive" |
| V | 16* | 163 | Psychological illness or psychosis: Anxiety, depression, epilepsy (valproate), confusion, borderline, lithium, memory problems, mood problems, dementia, seizure, psychotic, intellectual disability, BPD/BPAD (bipolar disorder/bipolar affective disorder), neuropathology/neuropathy |
| V | 17* | 163 | Alcohol abuse |
| IX | 18 | 189 | Cardiovascular, hypertension and hyperlipidemia: High blood pressure, high cholesterol, weight increase, chest pain, palpitations, on lipid lowering meds, tachycardia, CHD (congenital heart disease), pericardial effusion/oedema, arrhythmia, CRF (cardiac risk factors), CVA (cardiovascular accident), CHF (coronary heart failure), SOB (shortness of breath), review of vascular function/vascular disease, family history of CHD (chronic heart disease), familial hypercholesterolemia |
| X | 19 | 58 | Coughing, pneumonia, sore throat, emphysema, pneumothorax, restrictive lung disease |
| XI | 20 | 55 | GORD (gastro-oesophageal reflux disease) |
| XI | 21 | 248 | Gastrointestinal pain: ABDO Pain, "epigastric pain", nausea, vomiting, diarrhorea, colic, RUQ (right upper quadrant of the abdomen) pain, LUQ pain, RIF (right iliac fossa) pain, vesico colic fistula; AAA (abdominal aortic aneurysm), gastritis |
| XI | 22* | 53 | Gastrointestinal disorders: Crohn's disease, IBS (inflammatory bowel syndrome), coeliac disease, food [gluten] allergy, colitis |
| XI | 23 | 21 | Pancreatitis |
| XI | 24* | 517 | Liver diseases: Liver failure, raised/deranged LFTs, liver swelling |
| XI | 25 | 274 | Liver screen, paracetamol overdose |
| XII and XIII | 26* | 333 | Rheumatology: Arthralgia, arthritis, ankylosing spondylitis, scleroderma, sciatica, TKA/TKR (total knee replacement), THR (total hip replacement), myalgia, myositis/myolysis, gout, back pain, rash (including "urticaria", acne, bone problems (osteoporosis), psoriasis, other chronic inflammatory disorders |
| XIV | 27 | 258 | Kidney diseases: Renal failure (ESRF), CKD (chronic kidney disease), ADPKD (autosomal dominant polycystic kidney disease), glomerulonephritis, nephrotic syndrome, raised CREAT, BUN, proteinuria, haematuria, oedema |
| XIV | 28 | 114 | Reproductive systems: Amenorrhoea, gynecological problems, pelvic pain, PID, miscarriage, pap smear investigation, ovarian cysts, prostate problems, PSA, erectile dysfunction |
| XV | 29* | 3746 | Pregnancy: Pre- and post-pregnancy screening, RANTS (routine antenatal screen), EDC + ECD ("estimated date of confinement" + "estimated confinement date"), blood groups |
| XIV, XV and XVI | 30 | 370 | Fertility, fertility check, pre-pregnancy check, planning pregnancy, LMP (late menstrual period), infertility?, sub-fertility, IVF, PCOS (Polycystic Ovarian Syndrome), ovarian cysts |
| XX | 31* | 187 | IVDA (IV drug abuse) or "opiate dependence", "on methadone" |
| XXI | 32* | 287 | Screening: ANU medical school, clinical training + work/study/insurance/travel/health project associated screen, prescreen |
| XXI | 33 | 93 | Antiviral assessment |
| XXI | 34 | 98 | Check response to HAV or HBV vaccination, post vaccination, check immunity, immunisation |
| XXI | 35 | 340 | STD/STI screen, STRUT (screening of rectum, urethra and throat) |
| XXI | 36 | 527 | At risk behaviours: "contact", "exposure" with hepatitis/HIV carrier, professional (e.g., "needlestick injury"), OREE (occupational |
| | | | exposure), PEP (post-exposure prophylaxis), UPIC (unprotected intercourse) |
| - | 37* | 247 | Fatigue: Lethargy, tiredness, "feeling unwell", dizziness, collapse, malaise, "run down", syncope |
| - | 38 | 57 | Jaundice |
| - | 39 | 90 | Travel: overseas travel or from overseas (or migration to Australia from high risk area), post travel |
| - | 40 | 48 | Test at "patient's request" (PT request) |
| - | 41 | 102 | Pre-test fasting status: Fasting, non-fasting, "12-hour fast" |
| - | 42 | 98 | Admission: Acute case admission, new admission, new assessment, "for assessment", assess for Rx |
| - | 43 | 1417 | Surgery: Pre-op, post-op (all surgery including transplantation), pre-cardiac surgery, pre-CABG/CABG (coronary artery bypass grafting) |
| - | 44 | 95 | Organ donation: Renal/organ transplant (donor), other organ transplant, stem cell donor, white cell donor, egg donor, platelet donor |
| - | 45 | 5288 | No clinical notes |
| - | 46 | 872 | Non-specific notes: Clinical notes available, but incomplete, incomprehensible, non-specific comments ("unable to transcribe with confidence", "NIL", "unwell", "all ok (?)", "routine", "screening (unspecified)", "Rx monitoring", "follow up", "review", "FI (for investigation)" or could not decide disease state with certainty (e.g., APH = "acute poikilocapric hypoxia" or "acute preoperative hemodilution" or "acetylphenylhydrazine") |
| | | Total: 18624 | |

**Figure S1. Numerical example 1**: Decision matrix table for clinical note (Category 1: Hepatitis B positive) and reference standard (HBsAg)

|  |  | Status of person according to Reference (gold standard): HBsAg | | |
|---|---|---|---|---|
|  |  | Positive (IU >= 1.6) | Negative (IU < 1.6) | |
| Clinical note (test): category 1 | Hepatitis B positive | 69 (True positive) | 45 (False positive) | 114 |
|  | Hepatitis B negative | 8 (False negative) | 57 (True negative) | 65 |
|  | Total | 77 | 102 | 179 |

*note*: n = 179, is obtained from total n = 241 (minus data not available, n = 62) = 241 – 62 = 179

Sensitivity (Sn) = татTP / (TP + FN) = 69 / (69 + 8) = 0.90 (95% CI: 0.81-0.95)

Specificity (Sp) = TN / (TN + FP) = 57 / (57 + 45) = 0.56 (0.46-0.66)

Positive predictive value (PPV) = TP / (TP + FP) = 69 / (69 + 45) = 0.61 (0.55-0.66)

Negative predictive value (NPV) = TN / (TN + FN) = 57 / (57 + 8) = 0.87 (0.78-0.93)

Positive likelihood ratio (LR+) = Sn / (1 – Sp) = 0.90 / (1 – 0.56) = 2.05 (1.61-2.56)

Negative likelihood ratio (LR-) = (1 – Sn) / Sp = (1 – 0.90) / 0.56 = 0.18 (0.09-0.37)

**Numerical example 2**: **Decision matrix table for clinical note (Category 2: Hepatitis C positive) and reference standard (HCV)**

|  |  | Status of person according to Reference (gold standard): HCV | | |
|---|---|---|---|---|
|  |  | Positive (IU >= 1.0) | Negative (IU < 1.0) | |
| Clinical note (test): category 2 | Hepatitis C positive | 101 (True positive) | 38 (False positive) | 139 |
|  | Hepatitis C negative | 17 (False negative) | 10 (True negative) | 27 |
|  | Total | 118 | 48 | 166 |

*note*: n = 166, is obtained from total n = 327 (minus data not available, n = 161) = 327 – 161 = 166

Sensitivity (Sn) = TP / (TP + FN) = 101 / (101 + 17) = 0.86 (95% CI: 0.78-0.91)

Specificity (Sp) = TN / (TN + FP) = 10 / (10 + 38) = 0.21 (0.10-0.35)

Positive predictive value (PPV) = TP / (TP + FP) = 101 / (101 + 38) = 0.73 (0.69-0.76)

Negative predictive value (NPV) = TN / (TN + FN) = 10 / (10 + 17) = 0.37 (0.23-0.54)

Positive likelihood ratio (LR+) = Sn / (1 – Sp) = 0.86 / (1 – 0.21) = 1.08 (0.92-1.27)

Negative likelihood ratio (LR-) = (1 – Sn) / Sp = (1 – 0.86) / 0.21 = 0.67 (0.34-1.40)

**Figure S2. Prevalence of acute hepatitis B and C infection in Australia, from year 1990 to 2017.**

Source: Global Hepatitis Report. Geneva: World Health Organization. 2017.

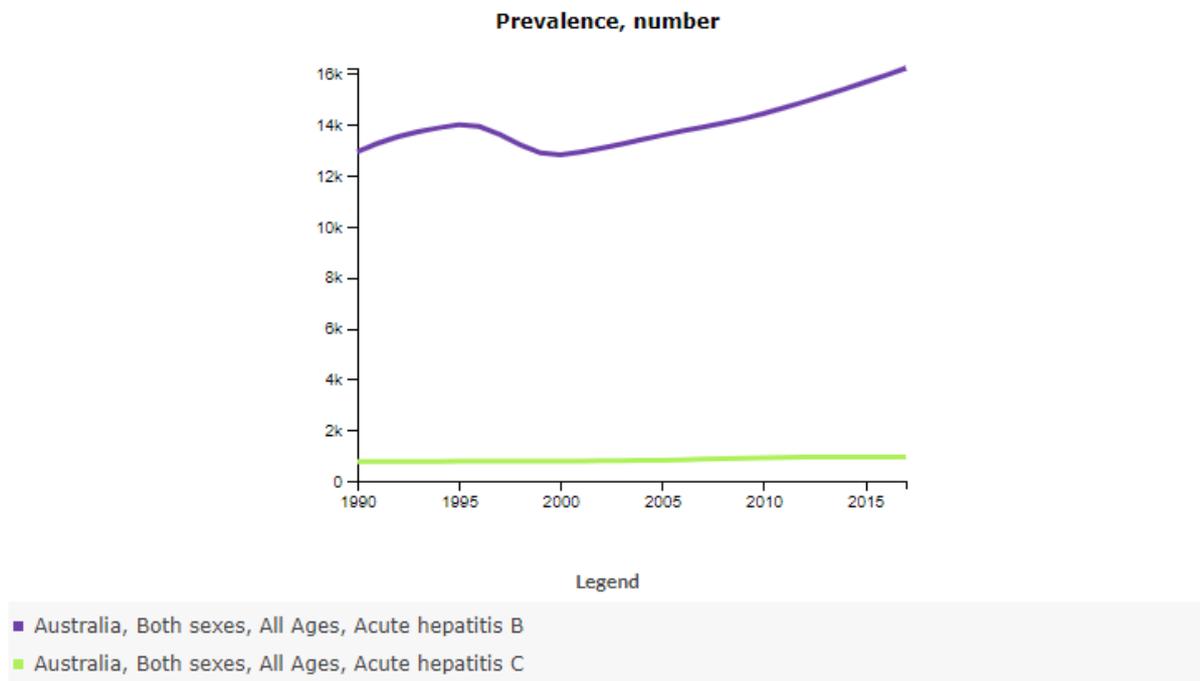